\documentclass[fp,st,twocolumn]{jpsj3}
\usepackage{txfonts}

\usepackage{color}
\usepackage{braket}
\usepackage{cite}

\newcommand{\pr}{%        
        ^\prime}
\newcommand{\ppr}{%
        ^{\prime\prime}}
\newcommand{\pppr}{%
        ^{\prime\prime\prime}}
\newcommand{\cc}{%
        c^\dag}    
\newcommand{\ca}{%
        c^{\phantom{\dag}}}  %small vector
\newcommand{\svek}{%
        \mathbf}

\newcommand{\new}[1] {{#1}}

\newcommand{\newblock} {}

\newcommand{\tsymb}{{\mathcal{T}}}
\newcommand{\torder}[1]{\left\langle  \tsymb \left[#1\right] \right\rangle}

\newcommand{\kvec}[1]{{\mathbf{#1}}}
\newcommand{\cvec}[1]{{\mathrm{#1}}}
\newcommand{\fref}[1]{Fig.~\ref{#1}}

\title{Towards {\em ab initio} calculations with the dynamical vertex approximation}

\author{A. Galler$^{1}$, J. Kaufmann$^{1}$, P. Gunacker$^{1}$,  P. Thunstr\"om$^{1,2}$, J. M. Tomczak$^{1}$, and K. Held$^{1}$}
\inst{$^{1}$  
Institute of Solid State Physics, TU Wien,  1040 Wien, Austria\\
$^2$Department of Physics and Astronomy, Materials Theory, Uppsala University, 751 20 Uppsala, Sweden} %\\

\abst{
While key effects of the many-body problem---such as Kondo and Mott physics---can be understood in terms of on-site correlations,
non-local fluctuations of charge, spin, and pairing amplitudes are at the heart of the most fascinating and unresolved phenomena in condensed matter physics.
Here,  we review recent progress in diagrammatic extensions to dynamical mean-field theory for {\it ab initio} materials calculations.
We first recapitulate the quantum field theoretical background 
behind the two-particle vertex. Next we discuss latest
algorithmic advances in quantum Monte Carlo simulations for calculating such two-particle quantities using worm sampling and vertex asymptotics, before  giving an introduction to the {\it ab initio} dynamical vertex approximation (AbinitioD$\Gamma$A).
Finally, we highlight the potential of AbinitioD$\Gamma$A by detailing results for the prototypical correlated metal SrVO$_3$.
}

\begin{document}
\maketitle

\section{Introduction}

It is particularly challenging to perform {\em ab initio} calculations of quantum materials if electronic correlations are strong.
In this case, the standard approach to materials calculations, density functional theory (DFT) \cite{Hohenberg1964,Kohn1965,Jones1989a} with exchange-correlation functionals
based on the local density approximation (LDA) or generalized gradient approximation (GGA) \cite{Perdew96},
is no longer reliable. Explicit many-body methods are needed.

A well-established route to this end is Hedin's \cite{Hedin1965} so-called 
{\it GW} approximation. Here, the self-energy is given by the product of the Green's function times a screened interaction: $\Sigma=G W$. This is the  Fock exchange diagram  but with
 $W$ screened  in the random phase approximation (RPA) instead of the bare interaction $V$. Hence, it is natural to expect it to improve on the exchange part of the exchange-correlation functional. Indeed this improved exchange allows {\it GW} to better predict 
the size of the band gaps in semiconductors \cite{Godby1988,Fuchs2007}. 
On top of this there are---because of the screening---genuine correlation effects such as quasiparticle renormalizations and finite life times \cite{Onida2002,Freysoldt_defects_RevModPhys_2014}. However, since it relies entirely on the RPA ladder and a first order expansion in the screened interaction,  {\it GW} is only appropriate for weakly correlated systems.

Strong electronic correlations are, on the other hand, the realm of dynamical mean field theory (DMFT) \cite{Metzner1989,Georges1992,Georges1996}. While DMFT does not cover all correlations, it includes the eminently important local correlations.
These do not only give rise to quasiparticle renormalizations and even metal-insulator transitions  \cite{Georges1996}, but also provide for a more reliable description of magnetism \cite{Jarrell1992,Held1998,Vollhardt98a} and kinks in the energy-momentum dispersion relation \cite{Byczuk2007,Held13}. Merging DFT and DMFT\cite{Anisimov1997,Lichtenstein1998} (for reviews, see Refs.~\citen{Kotliar2006,Held2007}) has been a big step forward for the calculation of strongly correlated materials.  Already the first DFT+DMFT calculations allowed for a better understanding of the Mott-Hubbard transition in V$_2$O$_3$\cite{Held2001a}, of $\delta$-Pu \cite{Savrasov2001}, the Ce volume collapse \cite{Held2001a}, and magnetism in Fe and Ni\cite{Lichtenstein2001}. More recently, among others, iron pnictides and their fluctuating magnetic moment \cite{Aichhorn2009,Hansmann2010,Ferber2012}, complex oxide heterostructures \cite{Hansmann2009,Hansmann2010b,Lechermann2014,Zhong2015,PhysRevB.92.041108}, nanoscopic structures \cite{Das2011,Valli2015}, thermoelectricity \cite{Arita2008,Wissgott2010,Tomczak12F}, spin-orbit interactions \cite{Martins2011,Arita2012}, and electronic entanglement \cite{Thunstrom2012}  have been addressed.

To include both, a better exchange and strong local correlations, DMFT has also been merged with {\it GW} \cite{Biermann2003}. The screening of the Coulomb interaction and the non-local {\it GW} self-energy has been found to be of importance in pnictides \cite{Werner2012,jmt_pnict,paris_sex},  intermetallics \cite{PhysRevB.82.085104}, cuprates \cite{PhysRevB.91.125142} and other transition metal oxides  \cite{Casula2012,Tomczak2012,Miyake13,0295-5075-99-6-67003,PhysRevB.93.235138}. Incorporating a {\it GW} correction into DFT+DMFT was, e.g., essential for describing  the red color of CeSF \cite{jmt_cesf}.

However, the non-local correlations of a  {\it GW}+DMFT calculation are  restricted to the RPA  screening, other non-local correlations are not taken into account. A prominent example of such  non-local correlations are spin fluctuations \cite{Moriya1985}, which are diverging in the vicinity of a magnetic phase transition and which can mediate high temperature superconductivity.\cite{Moriya2000,Scalapino12}

A way to address non-local correlations and to keep, at the same time, the local correlations of DMFT are cluster extensions of DMFT such as the dynamical cluster approximation (DCA) \cite{Hettler1998} and  cellular DMFT (CDMFT) \cite{Lichtenstein2000,Kotliar2001}.  Here, a cluster of sites, or a coarse-grained Brillouin zone is embedded in a non-interacting DMFT-like bath instead of a single site in DMFT. This way, non-local correlations within the cluster are accessible. These cluster extensions  helped establishing the presence of pseudogaps, superconductivity and their interplay %Jan: interplay refers to Fratino2016, i.e. Giovanni Sordi, which I would like to cite here.
in the two dimensional Hubbard model, see e.g.\ Refs.~\citen{Sakai2009,Sakai12,Gull2013,Fratino2016} and, for a review,  Ref.~\citen{Maier2005}.
For {\em ab initio} calculations with many orbitals on the other hand, the numerical effort increases too quickly with cluster size so that  cluster extension of DMFT have been hitherto restricted to two sites for materials calculations.\cite{Biermann2005,Lee2012}

In this respect, diagrammatic extensions of DMFT such as the dynamical vertex approximation (D$\Gamma$A) \cite{Kusunose2006,Toschi2007} are much more promising. 
In D$\Gamma$A,  and other related diagrammatic extensions \cite{Slezak2009,Rubtsov2008,Rohringer2013,Taranto2014,Li2015,Ayral2015}, first a local two-particle vertex is calculated. From this local building block, non-local Feynman diagrams are constructed through the Bethe-Salpeter  \cite{Toschi2007,Katanin2009} or through the parquet equations\cite{Held2014,Valli2015,Li2016}.
Short- and long-ranged correlations are treated on an equal footing. Among others, critical exponents could be calculated for the first time for the Hubbard model\cite{Rohringer2011,Hirschmeier2015} as well as the quantum critical point reached upon doping.\cite{Schaefer2016} In two dimensions, spin fluctuations suppress antiferromagnetic order and give rise to  pseudogap physics.\cite{Katanin2009,Jung2010,Otsuki2014} This yields a low temperature paramagnetic insulator at arbitrary small interaction\cite{Schaefer2015-2} and superconductivity.\cite{Otsuki2014,Kitatani2015} For a review see \cite{VertexRMP}.

For {\em ab initio} materials calculations, it has been suggested\cite{Toschi2011,Galler2016} to use as a starting vertex $\Gamma^{q}$ the bare non-local Coulomb interaction $V^{q}$ as well as all local vertex corrections  $\Gamma^{\rm loc}$, which also includes the local Coulomb (Hubbard) interaction $U$.
This way all {\it GW} diagrams and all local DMFT correlations are generated, but also non-local correlations beyond, such as spin fluctuations. Hence, this AbinitioD$\Gamma$A scheme includes more physics than {\it GW}+DMFT but requires the additional calculation of the local two-particle vertex and solving the Bethe-Salpeter equations in a Wannier basis. AbinitioD$\Gamma$A has been recently implemented and applied to study non-local correlations in SrVO$_3$.\cite{Galler2016}

In this paper we review the  AbinitioD$\Gamma$A method as well as the recent progress 
for calculating the multi-orbital local vertex.
In Section \ref{Sec:Ham}, we introduce the multi-orbital Hamiltonian that we want to solve and our notation regarding Green's functions and vertices.
Section
\ref{Sec:locG} is devoted to the continuous-time quantum Monte Carlo calculation (CT-QMC) of the local two-particle vertex, using vertex asymptotics\cite{Kaufmann2017} and worm sampling.\cite{Gunacker15}  In Section
\ref{Sec:abinitioDGA}, we recapitulate the  AbinitioD$\Gamma$A algorithm\cite{Galler2016,Note3} which starts with the local vertex plus the bare non-local interaction, constructs through the Bethe-Salpeter equation the non-local full vertex and through  the Schwinger-Dyson equation the non-local self-energy.
In Section  \ref{Sec:SVO}, we present the results obtained for SrVO$_3$ and (beyond
Ref.~\citen{Galler2016}) the difference in self-energy when using the vertex asymptotics in CT-QMC. Finally we provide a summary and outlook in Section \ref{Sec:conclusion}.

\section{Hamiltonian and formalism}
\label{Sec:Ham}

Our starting point is the following multi-orbital Hamiltonian of interacting electrons in a solid:

\begin{align}
H&= \sum_{\svek{k}lm\sigma}\epsilon_{\svek{k}lm}^{\phantom{\dag}}\cc_{\svek{k}m\sigma}\ca_{\svek{k}l\sigma}
\label{eq:Ham} \\
&+\sum_{\svek{k}\svek{k\pr}\svek{q}}\sum_{\substack{ll\pr mm\pr\\ \sigma\sigma\pr}} \left(U^{\phantom{\svek{q}}}_{lm\pr ml\pr}+V^{\svek{q}}_{lm\pr ml\pr}\right) \cc_{\svek{k\pr -q}m\pr \sigma} \cc_{\svek{k}l \sigma\pr} \ca_{\svek{k-q}m \sigma\pr}\ca_{\svek{k\pr}l\pr\sigma} \nonumber
\end{align}

Here  $\cc_{\svek{k}l\sigma}$ ($\ca_{\svek{k}l\sigma}$) creates (annihilates) an electron with momentum $\svek{k}$ and spin $\sigma$ in orbital $l$.
The first term in Eq.~(\ref{eq:Ham}), consisting of the one-particle dispersions, is usually obtained
from $GW$- or DFT-based methods. 
%The latter solve the solid's Schr{\"o}dinger equation in the Born-Oppenheimer approximation and employ an approximative mapping to an effective non-interacting system. % Both GW and DFT are already introduced in the introduction. 
After a physically relevant low-energy subset of the resulting
band-structure is identified, the hoppings are expressed in a suitable basis to yield the above amplitudes $\epsilon_{\svek{k}lm}$.
The second term in the Hamiltonian describes the interaction between the electrons.
The matrix elements have been separated into local (Hubbard/Hund-like) $U$ contributions and purely non-local interactions $V^{\svek{q}}$, i.e., $\sum_{\svek{q}}V^{\svek{q}}=0$.\cite{Note1}
Interaction strengths 
corresponding to the chosen orbital subset can be obtained from {\it ab initio} methods such as constrained DFT\cite{constrainedLDA} or constrained RPA\cite{Aryasetiawan2004}.
The former computes energy costs for the removal or addition of electrons, while the latter simulates the polarization of the other electrons in the solid that screen the bare Coulomb interaction.%

\begin{figure}
\vspace{1em}
\begin{center}
\includegraphics[width=8.6cm]{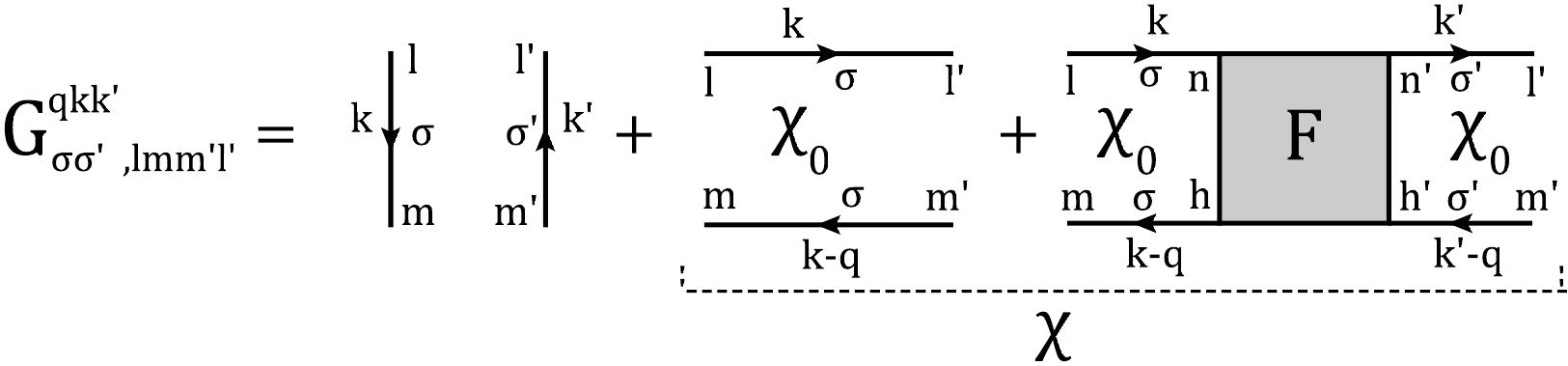}%
\end{center}
\caption{Decomposition of the two particle Green function ($G$) into disconnected parts and vertex ($F$) corrections. The last two terms constitute the generalized susceptibility $\chi=\chi_0+\chi_0F\chi_0$.}%
\label{fig:2pG}%
\end{figure}

Our general goal is to compute spectral and response properties for the many-body Hamiltonian defined in Eq.~\eqref{eq:Ham}.
This information can be extracted from the one-particle and two-particle Greens functions, which are respectively given by
\begin{align}
G^{\svek{k}}_{\sigma,lm}(\tau) & \equiv -\torder{ \ca_{\svek{k} l\sigma}(\tau) \cc_{\svek{k} m\sigma }(0) },\\
G^{\svek{q}\svek{k}\svek{k}\pr}_{\substack{ lmm\pr l\pr \\ \sigma\sigma\pr\sigma\ppr\sigma\pppr}}(\tau_1,\tau_2,\tau_3) & \equiv  
\label{eq:G}\\
\span\span\torder{  \ca_{\svek{k} l\sigma }(\tau_1) \cc_{\svek{k} - \svek{q}  m\sigma\pr }(\tau_2)  \ca_{\svek{k\pr}-\svek{q}  m\pr\sigma\ppr }(\tau_3) \cc_{\svek{k\pr} l\pr\sigma\pppr}(0) }\nonumber.
%G^{\svek{q}\svek{k}\svek{k}\pr}_{\substack{ lmm\pr l\pr \\ \sigma_1\sigma_2\sigma_3\sigma_4 }}(\tau_1,\tau_2,\tau_3) & \equiv & \\
%\span\span\torder{  \ca_{\svek{k} l\sigma_1 }(\tau_1) \cc_{\svek{k} - \svek{q}  m\sigma_2 }(\tau_2)  \ca_{\svek{k\pr}-\svek{q}  m\pr\sigma_3 }(\tau_3) \cc_{\svek{k\pr} l\pr\sigma_4}(0) }\nonumber.
\end{align}
where $\tsymb$ is the time-ordering operator, and brackets $\left\langle .\right\rangle$ denote the thermal expectation value in the action corresponding to the Hamiltonian in Eq.~\eqref{eq:Ham}.
Loosely speaking, the one-particle propagator describes the amplitude for the process in which one particle is added at time 0 to an ensemble of $N$ particles, the system is then let to evolve with $N+1$ particles, until one particle is again removed from the system at time $\tau$. The two particle propagator describes a similar process, except that two particles are added and removed at four different times. The propagators thus encode all possible scattering events on the one- and two-particle level.
If the Hamiltonian conserves spin (e.g., in the absence of spin-orbit coupling), there are only six non-vanishing spin combinations in the two-particle propagator:
\begin{align}
G^{\svek{q}\svek{k}\svek{k}\pr}_{\substack{ lmm\pr l\pr \\ \sigma\sigma\pr}}(\tau_1,\tau_2,\tau_3) & \equiv  G^{\svek{q}\svek{k}\svek{k}\pr}_{\substack{ lmm\pr l\pr \\ \sigma\sigma\sigma\pr\sigma\pr}}(\tau_1,\tau_2,\tau_3) \; ,
\label{eq:GF}\\
G^{\svek{q}\svek{k}\svek{k}\pr}_{\substack{ lmm\pr l\pr \\ \overline{\sigma\sigma\pr}}}(\tau_1,\tau_2,\tau_3) & \equiv  G^{\svek{q}\svek{k}\svek{k}\pr}_{\substack{ lmm\pr l\pr \\ \sigma\sigma\pr\sigma\pr\sigma}}(\tau_1,\tau_2,\tau_3) \; .
\end{align}
As depicted in Fig.~\ref{fig:2pG}, $G^{\cvec{qkk}}_{\substack{ lmm\pr l\pr \\ \sigma\sigma\pr}}$  can be decomposed as follows:
\begin{align}
G^{\cvec{q}\cvec{k}\cvec{k}\pr}_{\substack{ lmm\pr l\pr \\ \sigma\sigma\pr}}&=\beta\delta_{\cvec{q0}}G^{\cvec{k}}_{\sigma,lm}G^{\cvec{k\pr}}_{\sigma\pr, m\pr l\pr}
%-\beta\delta_{\sigma\sigma\pr}\delta_{\svek{kk\pr}}G^{k}_{\sigma,ll\pr}G^{k-q}_{\sigma, mm\pr}\nonumber\\
+\delta_{\sigma\sigma\pr}\chi_{0,\sigma, lmm\pr l\pr}^{\cvec{qkk\pr}}\nonumber \\
&+\sum_{nn\pr hh\pr}\chi_{0,\sigma, lmhn}^{\cvec{qkk}}F_{\substack{nhh\pr n\pr\\ \sigma\sigma\pr}}^{\cvec{qkk\pr}}\chi_{0,\sigma\pr, n\pr h\pr m\pr l\pr}^{\cvec{qk\pr k\pr}} \; .
\label{eq:G2}
\end{align}
\new{Here, $\beta$ is the inverse temperature, and  we have Fourier transformed to Matsubara frequencies according to}

\begin{equation}
G^{\cvec{q}\cvec{k}\cvec{k}\pr}_{\substack{ lmm\pr l\pr \\ \sigma\sigma\pr}} = \int_0^\beta d\tau_1 d\tau_2 d\tau_3 e^{i\nu\tau_1 - i(\nu-\omega)\tau_2 + i(\nu'-\omega)\tau_3} G^{\svek{q}\svek{k}\svek{k}\pr}_{\substack{ lmm\pr l\pr \\ \sigma\sigma\pr}}(\tau_1,\tau_2,\tau_3)
\label{Eq:Gchi}
\end{equation}
\new{and used the short-hand four-vector notation $\cvec{k}=(\svek{k},i\nu_n)$ and $\cvec{q}=(\svek{q},i\omega_n)$
for fermionic ($i\nu_n=(2n+1)\pi/\beta$) and bosonic ($i\omega_n=2n\pi/\beta$) variables, respectively.
This particular choice of the bosonic frequency $\omega$ corresponds to the particle-hole channel and is
coined particle-hole frequency notation. Additionally, one can define a particle-hole-transverse and a
particle-particle frequency notation as, e.\ g., in Refs.\ \cite{Rohringer2012,Wentzell2016}, but
in this case we use the convention described in \cite{Kaufmann2017}.
We have also introduced the bare two-particle propagator}
\begin{equation}
\chi_{0,\sigma, lmm\pr l\pr}^{\cvec{qkk\pr}}\equiv -\beta G_{\sigma, ll\pr}^{\cvec{k}}G_{\sigma, mm\pr}^{\cvec{k\pr-q}}\delta_{\cvec{kk\pr}} \; .
\label{eq:chi0}
\end{equation}
\new{and took in Eq.~\eqref{eq:G2} already into account that $\chi_{0}$ is diagonal in $\cvec{k}$,  $\cvec{k\pr}$}.

The first two terms in Eq.~\eqref{eq:G2} and Fig.~\ref{fig:2pG} represent the ``unconnected'' contributions
of the transversal and longitudinal  particle-hole channel (later referred to as ``$\overline{ph}$'' and ``$ph$''), respectively: the two particles propagate independently. 
The last, the ``connected'' term describes the actual scattering processes and consists of the full vertex $F$ with four "legs", which can be expressed by two bare two-particle propagators $\chi_0$, here written in the $ph$-notation.
The last two terms together, $\chi=\chi_0+\chi_0F\chi_0$ (see Fig.~\ref{fig:2pG}), form the generalized susceptibility 
from which, e.g., measurable charge- or spin-susceptibilities can be obtained by joining the external legs\cite{Park2011a,PhysRevB.92.205132}.

In case the system is paramagnetic and spin-orbit coupling can be neglected, there are only two independent
spin configurations for the two-particle quantities. It is particularly convenient to introduce the 
(d)ensity and (m)agnetic combinations, e.g., for $F$:
\begin{align}
F_{d} &\equiv F_{\uparrow\uparrow} + F_{\uparrow\downarrow},\\
F_{m} &\equiv F_{\uparrow\uparrow}-F_{\uparrow\downarrow}.
\end{align}
This way,  the spin indices are replaced by the label $r\in\{d,m\}$; for spin-independent quantities, such as $\chi_0$,  the $\sigma$-indices are henceforth omitted.
Scattering diagrams contained in the full vertex $F_r$ can further be divided into 
\begin{equation}
F_{r,lmm\pr l\pr}^{\cvec{qkk\pr}}=\Gamma_{r,lmm\pr l\pr}^{\cvec{qkk\pr}} + \phi_{r,lmm\pr l\pr}^{\cvec{qkk\pr}}
\label{eq:F1}
\end{equation}
where $\Gamma_r$ is the $ph$-irreducible vertex, which means that the diagrams contained in $\Gamma$  cannot be separated into two parts by cutting two horizontal propagator lines ($\chi_0$). The $\phi_r$ contains all the remaining diagrams, i.e., those that are $ph$-reducible and thus contains a section given by $\chi_0$. 
Therewith, all reducible terms can be constructed from the irreducible ones in $\Gamma_r$ via
\begin{equation}
%\phi^{qkk\pr}_{r,lmm\pr l\pr}=\sum_{nn\pr hh\pr \bar{k}k\ppr}\Gamma_{r,lmhn}^{qk\bar{k}}\chi_{0, nhh\pr n\pr}^{q\bar{k}k\ppr} F_{r,n\pr h\pr m\pr l\pr}^{qk\ppr k\pr}
\phi^{\cvec{qkk\pr}}_{r,lmm\pr l\pr}=\sum_{nn\pr hh\pr \cvec{k\ppr}}\Gamma_{r,lmhn}^{\cvec{qk{k\ppr}}}\chi_{0, nhh\pr n\pr}^{\cvec{q{k\ppr}k\ppr}} F_{r,n\pr h\pr m\pr l\pr}^{\cvec{qk\ppr k\pr}}
\label{eq:F2} \; .
\end{equation}
Together, Eqs.~\eqref{eq:F1}-\eqref{eq:F2} constitute the Bethe-Salpeter equation (BSE) for $F_r$ in the longitudinal particle-hole channel. Rewriting the BSE for generalized susceptibilities, $\chi=\chi_0+\chi_0\Gamma\chi$, shows that $\Gamma$ and the BSE are the two-particle analogue of the self-energy $\Sigma$ and the Dyson equation, $G=G_0+G_0\Sigma G$.

Instead of Eq. \eqref{eq:F1}, $F_r$ can equally be decomposed into a transversal particle-hole ($\overline{ph}$)-irreducible $\Gamma_{\overline{ph}}$ and a $\overline{ph}$-reducible
$\phi_{\overline{ph}}$.  
To distinguish $\Gamma$ and $\phi$ in different channels,
we use the index $\overline{ph}$ for the  $\overline{ph}$-channel
and no index for the here more widely used $ph$-channel.
Both channels  are  closely connected by the so-called crossing symmetry,
which for  $F$ reads:
\begin{align}
F_{\substack{lmm\pr l\pr \\ \sigma\sigma\pr}}^{\cvec{qkk\pr}} &= -F_{\substack{m\pr mll\pr \\ \overline{\sigma\pr \sigma}}}^{(\cvec{k\pr-k})(\cvec{k\pr-q})\cvec{k\pr}} = 
 \phantom{-}F_{\substack{m\pr l\pr lm \\ {\sigma\pr \sigma}}}^{(\cvec{-q})(\cvec{k\pr -q})(\cvec{k-q})}
% &=& -F_{\substack{ll\pr m\pr m \\ \overline{\sigma \sigma\pr}}}^{(k-k\pr)k(k-q)}\\
%&=& \phantom{-}F_{\substack{m\pr l\pr lm \\ {\sigma\pr \sigma}}}^{(-q)(k\pr -q)(k-q)}
\label{eq:CS}
\end{align}
The crossing symmetry transformations correspond to the relabelling of the in- and out-going external lines of the two-particle diagrams contained in $F$.

Given a irreducible vertex $\Gamma_r$ and the bare propagator $\chi_0$, the solution of the BSE \eqref{eq:F1}-\eqref {eq:F2} allows for the full determination of the two-particle Greens function defined in Eq.~\eqref{eq:G2}. The latter in turn gives access to diverse susceptibilities and spectroscopic observables.
In addition, via the Schwinger-Dyson equation of motion for the one-particle Greens function, the vertex $F$ also gives access to the self-energy $\Sigma$:
\begin{align}
\Sigma^{k}_{\sigma,mm\pr} &= \sum_{ll\pr\svek{k\pr}\sigma\pr}(U_{ml\pr l m\pr}+V^{\svek{q}=\svek{0}}_{ml\pr lm\pr})n^{\svek{k\pr}}_{\sigma\pr, ll\pr}\label{eq:EoM}\\
 &- \sum_{ll\pr\svek{q}}(U_{ml\pr lm\pr}+V^{\svek{q}}_{ml\pr lm\pr})n^{\svek{k}-\svek{q}}_{\sigma, ll\pr} \nonumber \\
&- \beta^{-1}\sum_{\substack{ll\pr nn\pr hh\pr\\ \cvec{q k\pr}}} \left( U_{mlhn}+V^{\svek{q}}_{mlhn}\right)
\chi_{0,nll\pr n\pr}^{\cvec{qk\pr k\pr}}F_{d,n\pr l\pr h\pr m\pr}^{\cvec{qk\pr k}} G^{\cvec{k-q}}_{hh\pr} \; .\nonumber
\end{align}
Here, the first two terms involving the density matrix $n^{\svek{k}}_{\sigma, ll\pr}$ are the (static) Hartree and Fock contributions, respectively. 
The crucial observation is that Eq.~\eqref{eq:EoM} allows for an alternative route to construct approximations for the self-energy:
Instead of selecting diagrams for the self-energy $\Sigma$ (or the one-particle propagator $G$), the diagrammatics can be lifted to
the two-particle level, i.e., approximations are made on the level of $F$ or $\Gamma$.
Previous works resorting to this philosophy in the context of many-body models include Refs.~\cite{Vilk1997,Takada2001,Kusunose2006,Toschi2007,Rubtsov2008,Katanin2009,Li2015}.
Here, we are going to describe AbinitioD$\Gamma$A\cite{Galler2016}, the dynamical vertex approximation for realistic electronic structure calculations.
Before that, we discuss however recent advances for calculating the local $F$ and $\Gamma$ by CT-QMC.

\section{CT-QMC calculation of the local vertex}
\label{Sec:locG}

%WIP
The two-particle Green's function, given by the  local version of  Eq.~\eqref{eq:G}, can be obtained
by solving a single-site impurity problem, e.g., at the end of a self-consistent DMFT calculation.
While a diversity of impurity solvers exists, in this context the CT-QMC algorithms 
are state-of-the-art due to their parameter robustness and numerical exactness.\cite{Rubtsov2005,Werner2006,Werner2006a} 
The latter is a result of the absence of any intrinsic discretization in the bath 
dispersion or the local impurity problem itself and (given enough computational resources) the convergence to the exact result.
 
In the strong coupling variant, the infinite series expansion of the partition function 
in terms of the hybridization function is sampled stochastically (CT-HYB). 
The resulting algorithm is especially suitable for multi-orbital systems at finite temperatures, 
which include local interactions beyond density-density interactions. 
In order to obtain the full local two-particle Green's function, 
we employ a worm sampling adaptation of CT-HYB.\cite{Gunacker15} 
That is, instead of extracting the Green's function estimates from the partition function series expansion by ``removing hybridization lines'', 
worm sampling additionally samples directly the Green's function series expansion. 
This allows for more flexibility in defining the Monte Carlo estimators, and hence to calculate all components of the two-particle Green's function.

Nevertheless, measuring the multi-orbital two-particle Green's function 
remains  challenging due to the vast amount of external degrees of freedom. 
Furthermore, CT-QMC algorithms naturally operate in the imaginary time basis, 
requiring an extra Fourier transform to Matsubara frequencies. 
Intrinsic statistical uncertainties in the high-frequency region of the two-particle 
quantities\cite{Hafermann2012,Gunacker2016} combined with the previously mentioned 
computational requirements limit us hence to small frequency box sizes of the two-particle Green's function.
 
In the following we show how to extract the asymptotic behavior of the local 
irreducible two-particle quantities.\cite{Kunes2011,Wentzell2016,Kaufmann2017}
Essentially this eliminates finite-size effects of the frequency box, and 
further improves the numerical results of the local vertex function.

We emphasize that CT-QMC algorithms make no distinction between reducible or irreducible 
and connected or disconnected diagrams. Instead, the native quantities supplied by 
the algorithm are one- and two-particle Green's functions.
It is thus necessary to subtract disconnected parts and amputate external propagators in order to obtain vertex functions.

\subsection{Vertex Asymptotics}
The asymptotical structure of two-particle quantities is usually considered at the level 
of the vertex functions $F$ or $\Gamma$, as certain frequency structures remain constant in the high-frequency limit.
Contrary, at the level of the two-particle Green's function 
any frequency structure approaches zero due to the decay %damping 
resulting from the external single-particle propagators.

A key observation is that the aforementioned asymptotical structure of the full vertex
consists of so-called Kernel functions\cite{Li2016} with a reduced dependence on only one or two Matsubara frequencies.
This directly leads to equal time components of the two-particle Green's function, 
accessible by CT-QMC algorithms.

\begin{figure}
\centering
\includegraphics[width=7.5cm]{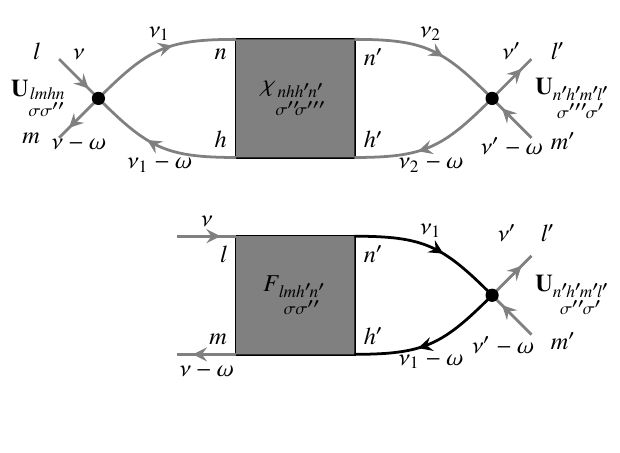}
\vspace{-3em}
\caption{\label{fig:eqT-GF}Single- and two-frequency vertex functions reducible in the particle hole channel. Grey lines only denote the 
incoming and outgoing frequencies and momenta,  whereas black lines denote internal propagators $G$. Please note that we have adjusted the index convention to the one in Ref.~\citen{Galler2016} which is slightly different from Refs.~\citen{Kaufmann2017,VertexRMP}.}
\end{figure}

 Let us start by motivating the asymptotical structure in terms of Feynman diagrams, cf.\ Refs.~\citen{Rohringer2012,Wentzell2016,VertexRMP}.
Diagrams which depend on a single bosonic frequency argument are essentially
two-particle diagrams $\chi$ (generalized susceptibilities), 
where the outer legs are contracted to equal-time pairs by two bare (static) interaction vertices 
in all possible combinations (upper row in Fig.~\ref{fig:eqT-GF}). 
This results in three single-frequency (sometimes referred to as two-legged) quantities, 
which are reducible in the particle-hole ($ph$), the particle-hole transversal ($\overline{ph}$) 
or the particle-particle ($pp$) channel. We refer to them as Kernel-1 functions, \cite{Li2016} which for the particle-hole channel read:
\begin{align}
K^{(1)\omega}_{\substack{ph, lmm\pr l\pr \\ \sigma\sigma\pr}} &= -\sum_{\substack{ nn\pr hh\pr \\ \sigma\ppr\sigma\pppr}} \mathbf{U}^{\phantom{\omega}}_{\substack{ lmhn \\ \sigma\sigma\ppr}} \chi_{\substack{ph, nhh\pr n\pr \\ \sigma\ppr\sigma\pppr}}^{\omega} \mathbf{U}^{\phantom{\omega}}_{\substack{ n\pr h\pr m\pr l\pr \\ \sigma\pppr\sigma\pr}}\label{eq:K1_1},
\end{align}
where $\mathbf{U}$ is the local Coulomb interaction written in form of a crossing-symmetric (i.e., Eq.~\ref{eq:CS}-fulfilling) vertex:  %(cf.\ Eq.~\ref{eqn:gammastart}).
\begin{align}
\mathbf{U}_{\substack{ lmm\pr l\pr \\ \sigma\sigma\pr}} &\equiv \beta^{-2} \Big( U_{lm\pr ml\pr} - \delta_{\sigma\sigma\pr} U_{ll\pr mm\pr} \Big).\label{symU}
\end{align}
Here and in the following,  we indicate quantities that explicitly fulfill the crossing relations by bold letters.
Also note that in the above equation, the full one-frequency susceptibilities $\chi_{ph}=\chi_{ph}^{\mathrm{con}} + \chi_0$ are used,
where the ``bubble'' term $\chi_0$ is given by summing over the fermionic frequencies in Eq.~\eqref{eq:chi0}.
%The indices $\sigma$ and $\sigma\pr$ encode the spin degrees of freedom, and $\bar{\sigma}$, $\bar{\sigma}\pr$ are their inverted values, e.g.\ $\sigma=\uparrow \Leftrightarrow \bar{\sigma}=\downarrow$.
As a by-product, the measurement of the single-frequency susceptibility in CT-QMC gives the physical local susceptibility. 

In a similar fashion, all diagrams which depend on one bosonic and one fermionic frequency argument 
are given by  diagrams
where only one pair of outer legs is contracted by the interaction vertex $\mathbf{U}$, 
while the other two legs are amputated (lower row in Fig.~\ref{fig:eqT-GF}).
This results in six two-frequency [sometimes referred to as three-legged (two fermionic, one bosonic)] quantities
or Kernel-2 functions\cite{Li2016}. These diagrams are reducible in different channels,
depending on which pair of outer legs is contracted.
Assuming time-reversal symmetry the number of diagrams is reduced from six to three. 
The Kernel-2 function that is reducible in the particle-hole channel, is given by
\begin{equation}
 K^{(2)\omega\nu}_{\substack{ph, lmm\pr l\pr \\ \sigma\sigma\pr}} = - \sum_{\substack{ \nu_1 n\pr h\pr \\ \sigma\ppr}}\,
  F^{\omega\nu\nu_1}_{\substack{ph, lmh\pr n\pr \\ \sigma\sigma\ppr}} \chi_{0, n\pr h\pr h\pr n\pr}^{\omega \nu_1\nu_1} \,
% G^{\nu_1}_{n'} G^{\nu_1-\omega}_{h'}\, 
  \mathbf{U}_{\substack{ n\pr h\pr m\pr l\pr \\ \sigma\ppr\sigma\pr}}
  -  K^{(1)\omega}_{\substack{ph, lmm\pr l\pr \\ \sigma\sigma\pr}},\label{eq:K2_1}
\end{equation}
where the Kernel-1 function is subtracted in order to avoid a double-counting of diagrams. Note that all these equations are local in Wannier space, i.e., we only have the frequency not the momentum indices.

Having established the reducible properties of the single- and two-frequency objects, 
the remaining task is to combine the kernel functions in order to approximate the full vertex function $F$.
It is thus helpful to decompose the local full vertex $F$ according 
to the (local)  Parquet equations \cite{DeDominicis1962,DeDominicis1964,Bickers2004}:
\begin{equation}
\label{eq:parquet} 
%F_{ijkl} = \Lambda_{ijkl} + \Phi^{ph}_{ijkl} + \Phi^{\overline{ph}}_{ijkl} + \Phi^{pp}_{ijkl},
F = \Lambda + \phi_{ph} + \phi_{\overline{ph}} + \phi_{pp} \; .
\end{equation}
Here $\Lambda$ is the fully two-particle irreducible vertex function, 
and $\phi_{ph},\phi_{\overline{ph}},\phi_{pp}$ are the two-particle reducible vertex 
in the particle-hole ($ph$), the particle-hole transversal ($\overline{ph}$) and the particle-particle ($pp$) channel [cf.\ Eq.~(\ref{eq:F1})].
For better readability, the spin and orbital indices are left out, and the channel dependent frequency arguments are also omitted at this point.
We are now able to construct the asymptotic form of the reducible vertices $\phi$ using:\cite{Wentzell2016}
\begin{equation}
%\Phi^{\text{asympt},{ph},\nu\nu'\omega}_{abcd} = K^{(1),{ph},\omega}_{abcd} + K^{(2),{ph},\nu\omega}_{abcd} + \overline{K}^{(2),{ph},\nu'\omega}_{abcd},
\phi^{\omega\nu\nu'}_{ph} = K^{(1)\omega}_{ph} + K^{(2)\omega\nu}_{ph} + \overline{K}^{(2)\omega\nu'}_{ph},
\end{equation}
where $\overline{K}^{(2)}$ is connected to ${K}^{(2)}$ by time-reversal symmetry.
The only remaining irreducible part contributing to the asymptotics of the full vertex function is the bare local interaction vertex $\mathbf{U}$ itself.
A more thorough derivation of the above equations is provided elsewhere.\cite{Kaufmann2017} 

\begin{figure}
\includegraphics[width=0.45\textwidth]{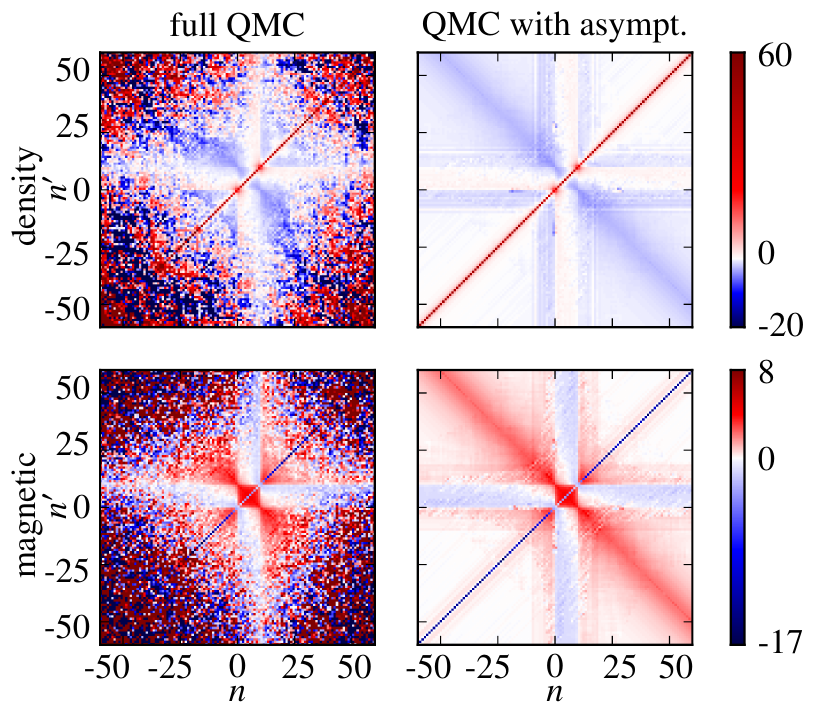}
\caption{\label{fig:f-svo}Full vertex $F^{d/m,\omega_{10}\nu\nu\pr}_{llll}$ with four equal orbital indices of SrVO$_3$ in the density (top) and magnetic channel (bottom) at inverse temperature $\beta=10$/eV.
  The left column shows the vertex as calculated directly from CT-QMC measurement. In the right column, 
  additionally the vertex asymptotics was employed. The constant background ($+U=5\,$eV in the density and $-U=-5\,$eV in the magnetic channel)
  was subtracted.}
\end{figure}

%The above procedure is necessary to extract the asymptotic form of the full vertex function $F$, including reducible components of each channel. %One could imagine other ways, e.g., using other model functions..
In \fref{fig:f-svo} we show the improvement introduced by the vertex asymptotics by comparing a slice of $F$ for the benchmark material SrVO$_3$ in the density and magnetic channel,
each with and without vertex asymptotics. Here, the two-particle $F$ is calculated after DFT+DMFT convergence,  see Section \ref{Sec:abinitioDGA} for further details of the calculation and for the strength of the Coulomb interactions.

When restricting ourselves to a subset of diagrams, as in the case of the two-particle irreducible vertex in the particle-hole channel $\Gamma_{ph}$,
the asymptotical structure is likewise reduced, because
the reducible diagrams of the particle-hole channel (i.e. $K^{(1)}_{ph}$ and $K^{(2)}_{ph}$) drop out of the description.
In the asymptotical region, the remaining Kernel-2 functions (i.e. $K^{(2)}_{\overline{ph}}$ and $K^{(2)}_{pp}$) eventually vanish, leading to the 
behavior shown in Ref.~\citen{Kunes2011}.

\begin{figure}
\includegraphics[width=0.45\textwidth]{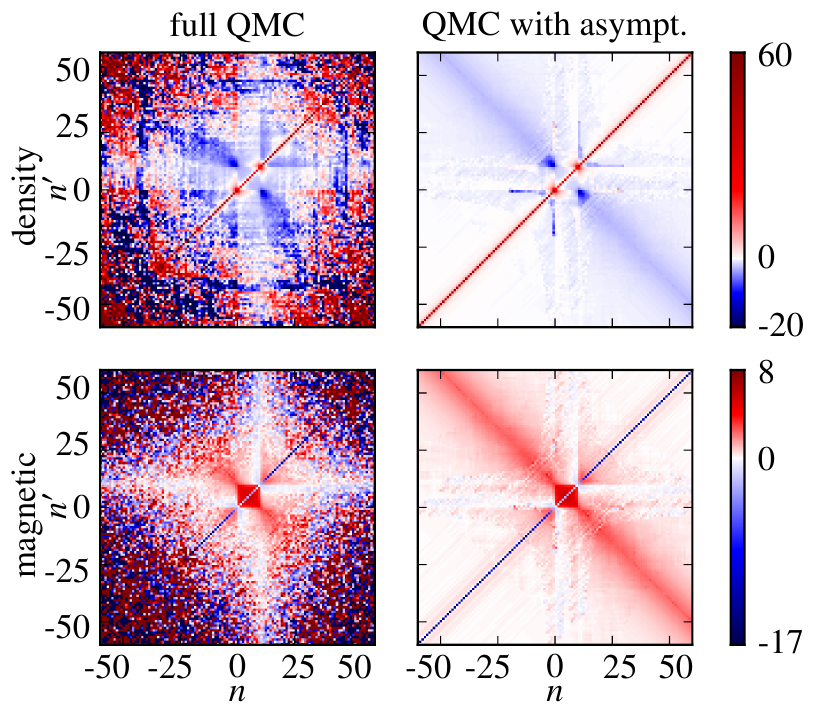}
\caption{\label{fig:gamma-svo}Same as Fig.~\ref{fig:f-svo}  but for the irreducible vertex $\Gamma^{d/m,\omega_{10}\nu\nu\pr}_{llll}$ in the $ph$-channel.
  The  left column  was calculated by inverting the BSE without asymptotics, whereas the results in the right column were obtained with the help of the asymptotic behavior (see text for details). }
\end{figure}

\fref{fig:gamma-svo} shows slices of the irreducible vertices of SrVO$_3$ in the density and magnetic channel  at fixed $\omega_{10}$.
The inner part was calculated from $F$ by inverting the respective BSE, using asymptotically enlarged susceptibilities.
For the outer part, the asymptotics procedure was applied directly  at the level of $\Gamma_{ph}$ as discussed in the previous paragraph. This eliminates a considerable part of the noise stemming from the matrix inversion.
Still, we notice that in the density channel it is numerically problematic to invert the BSE. 
Presumably, this is rooted in the vicinity of vertex divergencies  \cite{Schafer2013, Janis2014,Gunnarsson2016,Schaefer2016c,Ribic2016} in the phase diagram. 
Let us note that, to avoid these divergencies, the BSE is eventually reformulated in terms of the local $F$ instead of $\Gamma$ (see Ref.~\citen{Galler2016}), but for pedagogical reasons we focus here and in the following on the simpler  formulation in terms of  $\Gamma$.

We reiterate that the Kernel-1 and Kernel-2 functions are accessible in CT-HYB by measuring equal-time components of the two-particle Green's function.
For non-density-density interaction this requires a worm sampling algorithm. 
While it is still necessary to subtract any disconnected parts and amputate propagator lines,
the resulting approximation for the high-frequency region yields improved results. This way, the vertex function is not only defined on a larger frequency grid,
but also benefits from substantial improvements %enhancements 
in the intermediate frequency ranges. 

\begin{figure*}[tb]
\includegraphics[width=\textwidth]{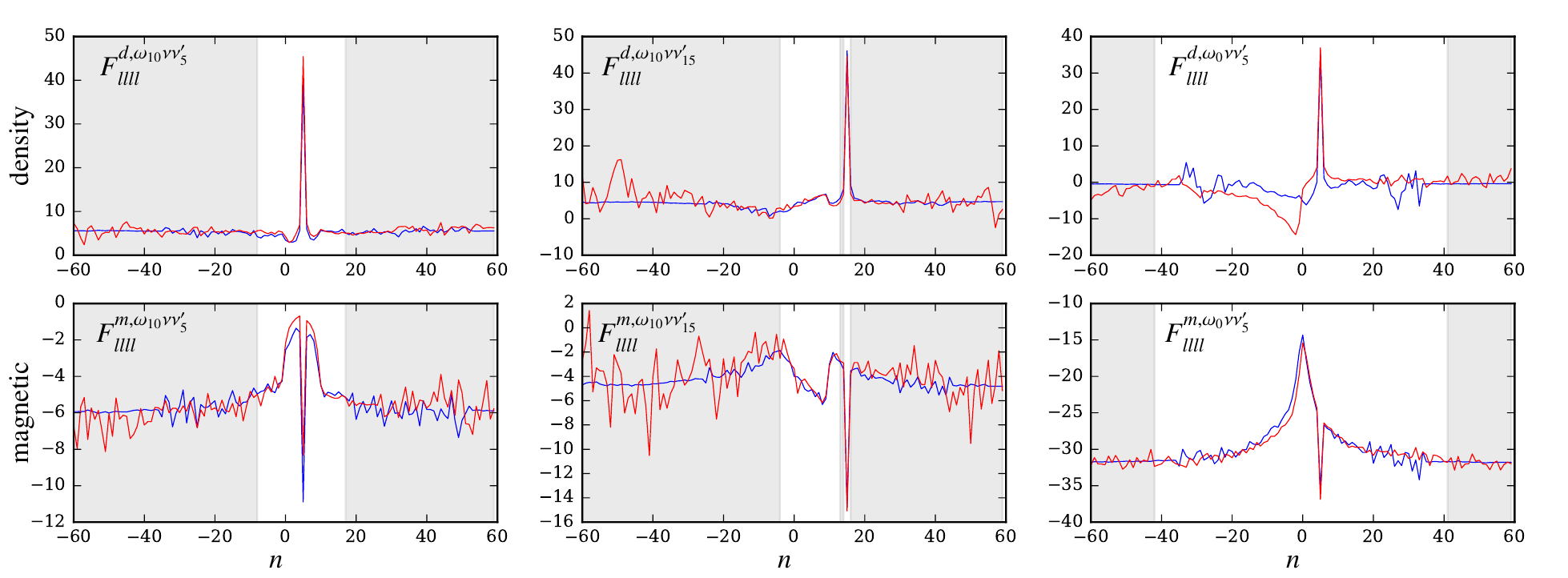}
\caption{\label{fig:f-slice-svo}
  \new{Full vertex $F^{d/m,\omega\nu\nu\pr}_{llll}$ as a function of $\nu$ for fixed $\omega$ and $\nu\pr$
  with four equal orbital indices of SrVO$_3$ in the density (top) and magnetic channel (bottom) at inverse temperature $\beta=10$/eV. 
  The red lines show the vertex, as conventionally calculated from CT-QMC data,
  whereas the blue lines show the purely asymptotic part of the vertex, and the
  shaded areas indicate where the asymptotic instead of the conventional CT-QMC data is used in the AbinitioD$\Gamma$A.}}
\end{figure*}

\new{
As the false color plots are not very quantitative,  we show in the first two columns of  
\fref{fig:f-slice-svo} two selected $\nu$-frequency cuts for fixed $\nu^\prime$ of the data displayed in  Fig.~\ref{fig:f-svo}
with the value of the full vertex on the  $y$-axis; % instead of a false color code; 
the last column shows two cuts for zero bosonic frequency, $\omega_0$. The shaded areas indicate where,  based on considerations for the atomic limit \cite{Kaufmann2017}, the asymptotic instead of the full QMC vertex is used in the AbinitioD$\Gamma$A.
This figure demonstrates that in the shaded region the asymptotic tail well represents the data without any noticeable off-set but with considerably less noise. 
}

\section{AbinitioD$\Gamma$A algorithm}
\label{Sec:abinitioDGA}
\begin{figure*}[tb]
\phantom{a}\hspace{-.8025cm}\begin{minipage}{15.83cm}
\begin{center}
\includegraphics[width=15.83cm]{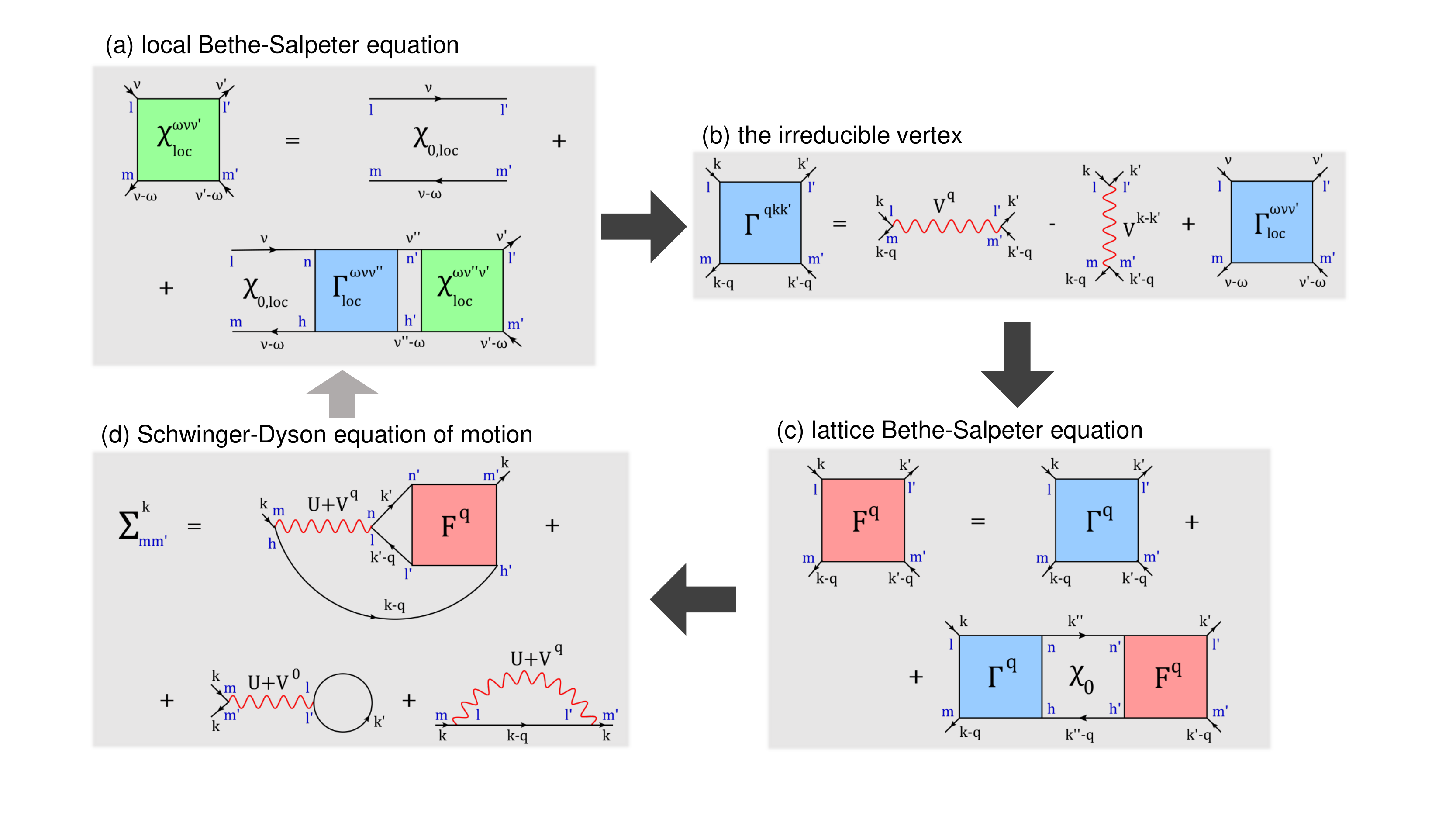}%
\end{center}
\end{minipage}\hspace{-.95cm}
\begin{minipage}{3.56cm}\vspace{-.9cm}
\caption{Flow diagram of the AbinitioD$\Gamma$A algorithm. (a) After DMFT convergence, the local BSE is solved to obtain the irreducible vertex $\Gamma$ using CT-QMC, see Section \ref{Sec:locG}. (b) The essential assumption  is approximating the  irreducible vertex by its local counterpart plus the bare non-local interaction $V^{\mathbf q}$, see Eq.~\eqref{eqn:gammastart}. (c) The lattice BSE [Eqs.~\eqref{eq:F1}-\eqref{eq:F2}] is employed to obtain the full vertex $F$ [Eq.~\eqref{eqn:ffinal} also includes the contributions from the 
 $\overline{ph}$-channel]. (d) Via the Schwinger-Dyson Eq.~\eqref{eq:final}, $F$ yields the  self-energy $\Sigma$.
With  $\Sigma$, a new Green's function and new impurity problem can be defined  in a self-consistent scheme. From Ref.~\citen{Galler2016}.
\label{fig:flow}}%
\end{minipage}
\end{figure*}

Let us assume that we have already obtained a self-consistent  DFT+DMFT (or $GW$+DMFT) solution. The first step of the AbinitioD$\Gamma$A algorithm
 [step (a) in Fig.~\ref{fig:flow}]  is to 
calculate, for the corresponding DMFT impurity model, the 
 \textit{local}  $ph$-irreducible vertex $\Gamma^{\omega\nu\nu\pr}_{\rm loc}$. 
For convenience we drop here and in the following text the index ``loc'', identifying local quantities by having frequency indices only.
How  $\Gamma^{\omega\nu\nu\pr}$ is calculated 
by means of CT-QMC has already been discussed in Section~\ref{Sec:locG}.

In AbinitioD$\Gamma$A  the $ph$-irreducible vertex is then approximated 
by this local $\Gamma^{\omega\nu\nu\pr}$ supplemented with the non-local Coulomb interaction ${\mathbf{V}}^{\kvec{q}\kvec{k}\kvec{k}'}$ [step (b) of the algorithm Fig.~\ref{fig:flow}]. This can be written in the form of a crossing-symmetric vertex: 
\begin{align}
\Gamma^{\cvec{q}\cvec{k}\cvec{k}'}_{\sigma\sigma',lmm\pr l\pr} & \equiv \Gamma^{\omega\nu\nu'}_{\sigma\sigma',lmm\pr l\pr} + {\mathbf{V}}^{\kvec{q}\kvec{k}\kvec{k}'}_{\sigma\sigma',lmm\pr l\pr},\label{eqn:gammastart}
\end{align}
where [cf.\ Eq.~(\ref{symU})]
\begin{align}
{\mathbf{V}}^{\kvec{q}\kvec{k}\kvec{k}'}_{\sigma\sigma',lmm\pr l\pr} & \equiv \beta^{-2} (V^{\kvec{q}}_{lm\pr ml\pr} - \delta_{\sigma\sigma'} V^{\kvec{k}'-\kvec{k}}_{mm\pr ll\pr}).\label{eqn:vqkk}
 \end{align}

Approximating the exact vertex by Eq.~\eqref{eqn:gammastart} is the essential approximation  of AbinitioD$\Gamma$A.
Since $\Gamma^{\omega\nu\nu\pr}$ already contains the local Coulomb interaction $U$ as its lowest-order contribution, Eq.~\eqref{eqn:gammastart} represents a natural extension of the local $\Gamma^{\omega\nu\nu\pr}$ to non-local interactions. 
That is, the  AbinitioD$\Gamma$A vertex $\Gamma^{\cvec{q}\cvec{k}\cvec{k}'}$ is made up from the local and non-local Coulomb interaction ($U$ and $V^{\kvec{q}}$) as well as all local vertex corrections.

The $ph$-irreducible vertex $\Gamma^{\cvec{q}\cvec{k}\cvec{k'}}$ defined in Eq.~\eqref{eqn:gammastart} is then used to calculate the full vertex function $F^{\cvec{q}\cvec{k}\cvec{k'}}$ through the BSE~\eqref{eq:F1}-\eqref{eq:F2} [as indicated in step (c) of Fig.~\ref{fig:flow}]. The latter can be considerably simplified if  $\Gamma^{\cvec{q}\cvec{k}\cvec{k}'}$ does not depend on the momenta $\kvec{k}$ and $\kvec{k}'$. Indeed, this dependence arises only from the second (crossed)
$V^{\kvec{k}'-\kvec{k}}$ term in Eq.\ \eqref{eqn:vqkk} which is, e.g., neglected in the $GW$ approximation\cite{Hedin1965}. If we follow the philosophy of $GW$ and neglect this term, Eq.~\eqref{eqn:gammastart} (now already in the two spin channels $r\in\{d,m\}$) reads    
\begin{equation}
\Gamma^{\cvec{q}\nu\nu'}_{r,lmm\pr l\pr}  = \Gamma^{\omega\nu\nu'}_{r,lmm\pr l\pr} + 2\beta^{-2} V^{\kvec{q}}_{lm\pr ml\pr} \delta_{r,d}.\label{eqn:gammavd} 
\end{equation}
With this simplification, the non-local  BSE [Eqs.~\eqref{eq:F1}-\eqref{eq:F2}] eventually becomes independent of $\kvec{k}$ and $\kvec{k}'$ and reads
\begin{align}
F^{\cvec{q}\nu\nu'}_{r,lmm\pr l\pr}  & = \Gamma^{\cvec{q}\nu\nu'}_{r,lmm\pr l\pr} + \phi^{\cvec{q}\nu\nu'}_{r,lmm\pr l\pr}\label{eqn:fqnunu}\\
\phi^{\cvec{q}\nu\nu'}_{r,lmm\pr l\pr} & = \sum_{\substack{nn\pr hh\pr \\ \nu''}}\Gamma^{\cvec{q}\nu\nu''}_{r,lmhn} \chi^{\cvec{q}\nu''\nu''}_{0,nhh\pr n\pr} F^{\cvec{q}\nu''\nu'}_{r,n\pr h\pr m\pr l\pr},\label{eqn:phinonloc}
\end{align}
with 
\begin{equation}
\chi^{\cvec{q}\nu\nu}_{0,lmm\pr l\pr}  = \sum_{\kvec{k}} \chi^{\cvec{q}\cvec{k}\cvec{k}}_{0,lmm\pr l\pr} = -\beta \sum_{\kvec{k}}G^{\cvec{k}}_{ll\pr}G^{\cvec{k}-\cvec{q}}_{m\pr m}. \label{eq:bubble_q}
\end{equation}

The full vertex function $F^{\cvec{q}\nu\nu'}$ constructed in this way contains non-local diagrams only in the $ph$-channel. However, in D$\Gamma$A we consider also the corresponding non-local diagrams in the $\overline{ph}$-channel. The latter do not need to be constructed explicitly, but can be obtained through symmetry considerations, namely by the crossing symmetries of Eq.~\eqref{eq:CS}. 
Adding the $ph$ and  $\overline{ph}$-channel and subtracting any double counted term yields---after some algebra as discussed in  Ref.~\citen{Galler2016}---the full AbinitioD$\Gamma$A vertex:
\begin{align}
{\mathbf{F}}^{\cvec{q}\cvec{k}\cvec{k}'}_{d,lmm\pr l\pr} = & F^{\omega\nu\nu'}_{d,lmm\pr l\pr} + F^{nl,\cvec{q}\nu\nu'}_{d,lmm\pr l\pr} - \frac{1}{2} F^{nl,(\cvec{k}'-\cvec{k})(\nu'-\omega)\nu'}_{d,m\pr mll\pr} \nonumber\\
&- \frac{3}{2}F^{nl,(\cvec{k}'-\cvec{k})(\nu'-\omega)\nu'}_{m,m\pr mll\pr}.\label{eqn:ffinal}
\end{align}
Here, the non-local vertex $F^{nl}$ is defined as 
\begin{equation}
F^{nl,\cvec{q}\nu\nu'}_{r,lmm\pr l\pr} \equiv F^{\cvec{q}\nu\nu'}_{r,lmm\pr l\pr} - F^{\omega\nu\nu'}_{r,lmm\pr l\pr}\label{eqn:fnl}  \; 
\end{equation} 
with $F^{\cvec{q}\nu\nu'}$ calculated from $\Gamma^{\cvec{q}\nu\nu''}$ through the BSE~\eqref{eqn:fqnunu}-\eqref{eqn:phinonloc}.

The next step  [step (d) in Fig.~\ref{fig:flow}] is to use
the full vertex ${\mathbf{F}}$ of Eq.~\eqref{eqn:ffinal} in the Schwinger-Dyson equation of motion~\eqref{eq:EoM} to obtain the AbinitioD$\Gamma$A self-energy. 
Since in the BSE~\eqref{eqn:fqnunu}-\eqref{eqn:phinonloc} we included  ${V}^\kvec{q}$ through $\Gamma^{\cvec{q\nu\nu'}}$ but not ${V}^{\kvec{k}'-\kvec{k}}$, it is consistent to also neglect corresponding  terms for the self-energy in the  Schwinger-Dyson equation. We can also explicitly identify the contribution $\Sigma^{\nu}_{{\mathrm{DMFT}}}$ that corresponds to the DMFT solution. After some algebra (see Ref.~\citen{Galler2016} for details),  the final expression for the non-local AbinitioD$\Gamma$A self-energy reads
\begin{align}
\label{eq:final}
\Sigma^{{\mathrm{D}}\Gamma{\mathrm{A}~}} = & \Sigma^{\nu}_{{\mathrm{DMFT}}}-\beta^{-1} \sum_{\cvec{q}\nu'} \! U \chi^{nl,\cvec{q}\nu'\nu'}_{0} F^{\omega\nu'\nu}_{d} G^{\cvec{k}-\cvec{q}}  \nonumber \\
- & \beta^{-1} \sum_{\cvec{q}\nu'} \! {V}^{\kvec{q}} \chi^{\cvec{q}\nu'\nu'}_{0} F^{\omega\nu'\nu}_{d} G^{\cvec{k}-\cvec{q}} \nonumber \\
- & \beta^{-1} \sum_{\cvec{q}\nu'} \! \Big( U + {V}^{\kvec{q}}\Big) \chi^{\cvec{q}\nu'\nu'}_{0} F^{nl,\cvec{q}\nu'\nu}_{d} G^{\cvec{k}-\cvec{q}} \nonumber \\
+ & \beta^{-1} \sum_{\cvec{q}\nu'} \! \tilde{U} \chi^{\cvec{q}\nu'\nu'}_{0} \Big(\frac{1}{2} F^{nl,\cvec{q}\nu'\nu}_{d} + \frac{3}{2} F^{nl,\cvec{q}\nu'\nu}_{m} \Big)G^{\cvec{k}-\cvec{q}}.
\end{align}
Here $\tilde{U}_{lm\pr l\pr m} = U_{lm\pr ml\pr}$; % and similarly for $V$;  
$\chi^{nl,\cvec{q}\nu\nu}_{0} \equiv \chi^{\cvec{q}\nu\nu}_{0} - \chi^{\omega\nu\nu}_{0}$;  and orbital indices have been suppressed for clarity.

This self-energy contains all DMFT and $GW$ contributions as well as non-local correlations beyond (e.g., spin fluctuations). It is constructed from the underlying AbinitioD$\Gamma$A approximation of considering only local vertex corrections. From the self-energy, we can also calculate the Green's function via the Dyson equation. This in turn allows us to define a new impurity problem in a self-consistent scheme, as indicated in Fig.~\ref{fig:flow}. Let us now turn to the results obtained by AbinitioD$\Gamma$A, so far  without  self consistency and  without ${V}^{\kvec{q}}$.

\paragraph{Numerical effort}
\new{
The numerical bottleneck is, in the case of SrVO$_3$, the computation of the local vertex in CT-QMC, 
which roughly scales as $\beta^5\#o^4$. Here, $\#o$ is the number of orbitals and there is a large prefactor because of the Monte-Carlo sampling.   Calculating the vertex for SrVO$_3$ with $\#o=3$, $\#\omega=120$ Matsubara frequencies and $\beta=10\,$eV$^{-1}$ took 150000 core h (Intel Xeon E5-2650v2, 2.6 GHz, 16 cores per node). Calculating the asymptotic form of the vertex instead only scales as $\beta^4(\#o)^4$. This shows that a room temperature calculation of the full QMC vertex with an equally fine $\#\omega$ grid is hardly feasible with present-day computational resources, but should be possible when using the vertex asymptotics in a large part of the frequency box.
}

\new{
As for the subsequent  AbinitioD$\Gamma$A calculation, the Bethe-Salpeter equation 
has to be solved independently for all  $\cvec{q}$-points, i.e.,
for $\#q$  momentum-points and  $\#\omega \sim \beta$ (bosonic) Matsubara frequencies. For every such  $\kvec{q}$-point, a matrix of dimensions $\#\omega (\#o)^2$ is inverted. Altogether the numerical effort hence roughly (depending on the matrix inversion algorithm) scales as $(\#q\#\omega)  (\#\omega\#o^2)^{2.5}$. Solving the  Bethe-Salpeter equation obviously becomes the computational bottleneck if many orbitals are included and the temperature is not too low.
}

\section{Results for SrVO$_3$}
\label{Sec:SVO}

\subsection{The material}

The reviewed AbinitioD$\Gamma$A formalism has the potential to clarify the microscopic origins
of the non-local fluctuations associated, e.g., with (quantum) criticality. However, as a first application it is better to consider a prototypical system that has been well-studied with previous approaches. In the field of correlated
electrons the most popular benchmark material is SrVO$_3$\cite{PhysRevB.77.205112,Casula2012,Kazuma16,PhysRevB.94.155131}.
In this perfectly cubic perovskite oxide, the octahedral coordination of vanadium atoms splits their 3$d$
orbitals into a high-lying $e_g$ doublet and the lower, triply degenerate $t_{2g}$ orbitals.
Due to this crystal-field splitting, the nominal electronic configuration of vanadium is $t_{2g}^1e_g^0$.

Experimentally, diverse manifestations of many-body effects have been established in SrVO$_3$:
Photoemission spectroscopy\cite{Sekiyama2004} and specific heat measurements\cite{PhysRevB.58.4372} both find a mass enhancement of a factor of two
compared to band-theory, while effective masses extracted from optical spectroscopy are even slightly larger\cite{PhysRevB.58.4384}.
A closer inspection yields a kink in the self-energy and hence the energy-momentum dispersion \cite{Nekrasov05a,Aizaki12}. Concomitant with the band-width narrowing, the one-particle spectrum exhibits a satellite feature below
the quasi-particle peak\cite{PhysRevB.52.13711,Sekiyama2004,PhysRevB.80.235104,Yoshida2016}. 
This satellite has been interpreted as the lower Hubbard band, i.e., the remnant of the atomic-like $t_{1g}^1\rightarrow d^0$ multiplet
transition of a single vanadium atom in said crystal-field environment.

The effort to develop electronic structure methods capable of quantitative predictions for correlated materials
is fueled by the rich multitude of potential applications of these materials: In the case of SrVO$_3$ these include its potential use as
transparent conductor\cite{Zhang2016}, electrode material\cite{ADMA:ADMA201300900}, or Mott transistor\cite{Zhong2015}.

While initially thought to be well-described by now standard (back then: pioneering) DFT+DMFT calculations\cite{Sekiyama2004,Pavarini04,Nekrasov05a},
new aspects have been unraveled by recent advancements in, both, many-body electronic structure theory\cite{Tomczak2012,PhysRevB.88.235110,Tomczak2014,Boehnke2016},
and experimental techniques \cite{PhysRevB.94.241110,FujiPC}:
Calculations taking into account the dynamical nature of screening by allowing for retarded interactions
suggest that a substantial part of the mass renormalization in SrVO$_3$ is driven by plasmon excitations
\cite{Casula2012,0295-5075-99-6-67003,Tomczak2012,PhysRevB.88.235110,Tomczak2014,Boehnke2016}. Further, $V^q$-mediated screened exchange contributions to the self-energy
were found to be important \cite{Tomczak2012,Miyake13,Tomczak2014,0295-5075-108-5-57003,Boehnke2016}, and shown to compete with the renormalization of the quasi-particle weight \cite{jmt_pnict,Tomczak2014}.
These effects beyond DMFT were predicted to be most pronounced for unoccupied states\cite{Tomczak2014}, in quantitative congruence with subsequent experiments\cite{FujiPC}.
Finally, it was shown recently that the intensity of the occupied Hubbard satellite is sizably enhanced by the presence of oxygen vacancies\cite{PhysRevB.94.241110}. Whether this is quantitatively  compatible with the reduced Hubbard correlations evidenced in recent {\it GW}+DMFT calculations\cite{Tomczak2012,Tomczak2014,Boehnke2016} needs to be shown.
For details regarding these aspects, we refer the reader to the recent reviews Refs.~\citen{0953-8984-26-17-173202,0953-8984-28-38-383001,Tomczak2017}.

\subsection{AbinitioD$\Gamma$A for SrVO$_3$}

While the AbinitioD$\Gamma$A approach can in principle address all of the above mentioned effects beyond DMFT---$V^q$-mediated screened exchange, plasmon satellites, or oxygen vacancies---we
here focus on a single issue: How strong are non-local renormalizations in SrVO$_3$ when solving a three-band Hubbard model for the low-energy $t_{2g}$ orbitals of SrVO$_3$  with a static and purely local interaction?
We are thus using a setup as in standard DFT+DMFT calculations, but we relax the approximation of having only a local self-energy.

\paragraph{Prelude.}
The AbinitioD$\Gamma$A calculations were performed using the local vertex functions defined in Section \ref{Sec:locG}, computed for an inverse temperature $\beta=10$/eV, a Hubbard intra-orbital interaction $U=5.0$eV, and a Hund's exchange $J=0.75$eV. The AbinitioD$\Gamma$A self-energy, defined in Eq.~\eqref{eq:final}, was computed using the formalism of Section \ref{Sec:abinitioDGA} (with $V^q=0$).

\begin{figure}[tb]%
\includegraphics[width=9cm]{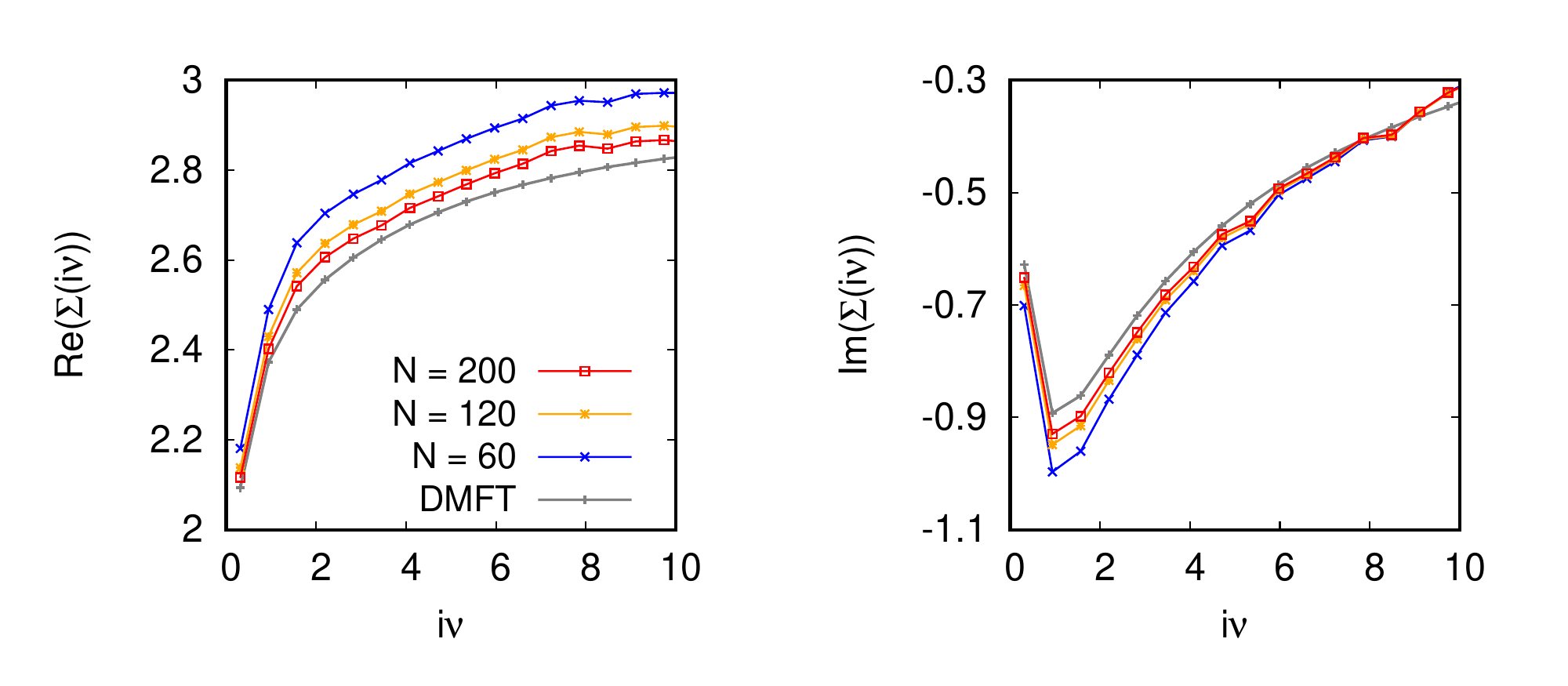}%
\caption{Comparison of the DMFT self-energy with the in principle equivalent contribution $\Sigma^\nu_{{\mathrm{DMFT}}}=-\beta^{-1}\sum_{\cvec{q}\nu\pr}U\chi_0^{\omega\nu\pr\nu\pr}F_d^{\omega\nu\pr\nu}G^{\cvec{k-q}}$
 obtained from the equation of motion [Eq.~\eqref{eq:final}]. Shown are the real (left) and imaginary (right) parts of the self-energy for different sizes $N$ of the frequency boxes $[-i\nu_N:+i\nu_N]$, $[-i\omega_N:+i\omega_N]$ used in the calculation.
}%
\label{fig:localS}%
\end{figure}

\begin{figure*}[tb]%
\begin{minipage}{13.cm}
\begin{center}
\includegraphics[width=\textwidth]{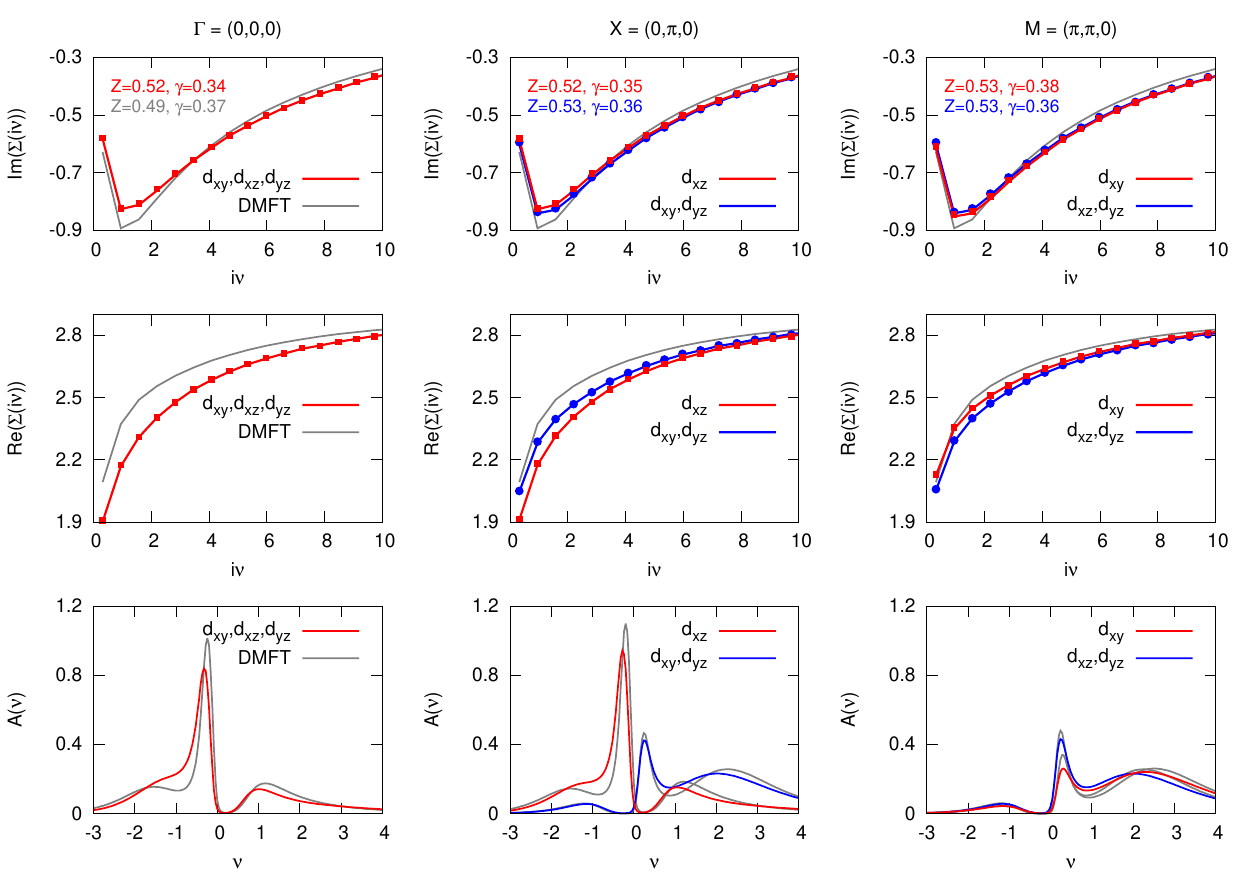}%
\end{center}
\end{minipage}\hfill
\begin{minipage}{3.8cm}
\caption{AbinitioD$\Gamma$A results for the self-energy (top row: imaginary part, middle row: real part) and real-frequency spectral functions (bottom row) for selected $\mathbf{k}$-points in the Brillouin zone
($\Gamma$-point: left column, $X$-point: middle column, $M$-point: right column).
In the two top panels, symbols indicate results for a small frequency box (N=60) and using no vertex asymptotics, while lines depict results for the largest box (N=200) that also makes use of the vertex asymptotics:
Both results are on top of each other.
\label{fig:ADGAresults}%
}%
\vfill
\end{minipage}
\end{figure*}

First, we discuss the convergence of the results with respect to the frequency box. While the techniques presented in Section \ref{Sec:locG} allow us to compute the local vertex
for arbitrarily large frequencies, the scaling of the AbinitioD$\Gamma$A formalism may cause both the memory consumption $\sim \#\omega^3$ ($\#\omega$: number of Matsubara frequencies) and the computing time  $\sim \#\omega^{3.4}$ to become prohibitively expensive.
In order to obtain fully converged results for the self-energy at affordable costs, we have separated out in Eq.~\eqref{eq:final} the term $\Sigma^\nu_{{\mathrm{DMFT}}}=-\beta^{-1}\sum_{\cvec{q}\nu\pr}U\chi_0^{\omega\nu\pr\nu\pr}F_d^{\omega\nu\pr\nu}G^{\cvec{k-q}}$. 
This quantity which should, in principle, equal the initial DMFT result is shown in Fig.~\ref{fig:localS} for different sizes of the frequency box.
While a clear tendency towards the correct DMFT result is seen, convergence is, however, not yet reached even for boxes extending to $\pm 1240/\beta$.
In our implementation we therefore substitute this term with the numerically exact DMFT result.  
The rationale behind this procedure is that the remaining $U$-terms in the AbinitioD$\Gamma$A self-energy either depend on the fully non-local free susceptibility $\chi_0^{nl,\cvec{q}\nu\nu}$ or the
non-local full vertex $F_d^{nl,\cvec{q}\nu\nu\pr}$, which both decay faster with frequency than their local counterparts by at least one power.
As a result, the AbinitioD$\Gamma$A self-energy is indeed quasi-independent of the size of the frequency box: The two top panels of Fig.~\ref{fig:ADGAresults} display, respectively, the imaginary and real parts of the
self-energy. The symbols indicate results for the smallest frequency box (N=60, as used in Ref.~\citen{Galler2016} without vertex asymptotics), while the lines depict the self-energy for the largest box (N=200) where the vertex asymptotics has been used.
This fast convergence with frequency suggests that the approach can easily be  applied to more complex systems as well as to lower temperatures. The high frequency asymptotics of the vertex, which was so important for calculating the vertex in a larger box at a lower noise level (see Section \ref{Sec:locG}), and is instrumental for obtaining reliable susceptibilities (see Ref.~\citen{Kaufmann2017}), does not reflect in large changes of the AbinitioD$\Gamma$A self-energy.

\paragraph{Results.}

We now discuss the effects in the AbinitioD$\Gamma$A self-energy---displayed in Fig.~\ref{fig:ADGAresults} for three representative $\mathbf{k}$-points---that are beyond DMFT.
At low energies the imaginary part of the self-energy on the Matsubara axis is---for all orbitals and $\mathbf{k}$-points---slightly smaller than in DMFT.
As a result the scattering rate $\gamma=-\Im\Sigma(i\nu\rightarrow 0)$ very slightly decreases with respect to DMFT, while the quasi-particle weight $Z_\mathbf{k}$ increases.
Besides this overall effect, the momentum dependence of $\Im\Sigma$ is small. Indeed $Z_\mathbf{k}$ varies by less than 2\% within the Brillouin zone.\cite{Note2}

The real-part of the self-energy shows larger deviations from the DMFT result. 
Indeed, at low energies the difference between AbinitioD$\Gamma$A and DMFT reaches 200meV, which amounts to about 7\% of the Hartree term.
Further, in DMFT the $d_{xy}$, $d_{xz}$, and $d_{yz}$ self-energy components were identical due to their local degeneracy.
As a consequence of the lifting of the momentum-independence, the self-energy can now also acquire an orbital dependence. The splitting between non-equivalent components
at the $X$ and $M$-point of the Brillouin zone is displayed in the middle and right panel of Fig.~\ref{fig:ADGAresults}. The orbital splitting at a given $\mathbf{k}$-point reaches up to 100meV.

Finally, we present  the spectral functions obtained from the AbinitioD$\Gamma$A approach in the bottom panels in Fig.~\ref{fig:ADGAresults}. Using a maximum entropy algorithm, we continue the Matsubara Green's function to the real-frequency axis. From the discussion of the self-energy, we expect differences to DMFT results of the order of 200meV,
with the tendency to reduce signatures of correlation effects.
This is indeed confirmed: Owing to the larger $Z$-factor as well as shifts mediated by $\Re\Sigma$ (cf.\ discussion below), the quasi-particle peaks move slightly away from the Fermi level
while Hubbard bands move towards it.

\paragraph{Discussion.}

It is instructive to analyze the structure of the AbinitioD$\Gamma$A self-energy in more detail.
The momentum and orbital dependence in the real-part of the self-energy join forces to push the occupied and unoccupied states away from each other:
At the $X$-point (middle column in Fig.~\ref{fig:ADGAresults}) the mostly occupied $d_{xz}$-states are further pushed away from the almost empty $d_{xy}$, $d_{yz}$-states, since $\Re\Sigma_{xz}(X,0)<\Re\Sigma_{xy,yz}(X,0)$.
In the same way, $\Re\Sigma_{xy,xz,yz}(\Gamma,0)<\Re\Sigma_{xy}(M,0)<\Re\Sigma_{xz,yz}(M,0)$. That is, 
at the $\Gamma$-point (where the quasi-particle peaks of all orbitals are below the Fermi level) all states are pushed down relative to all components at the $M$-point (where the quasi-particle peaks are above $E_F$), and more so the higher the initial DFT band was [$\epsilon_{xy}(M)<\epsilon_{xz,yz}(M)$]. 
In all, these effects of the non-degeneracy of the self-energy result in a band-width widening, or, equivalently speaking, in a reduction of effective masses.
We stress that this effect is distinct from the (often scissors-like) band-gap widening provided by hybrid functionals or {\it GW} calculations. In the latter, the effect arises from
exchange contributions that are mediated by non-local interactions $V^q$ (that are absent in the current calculation). 
For a more detailed discussion see Ref.~\citen{Galler2016}.

From a low-frequency expansion of the self-energy on the real-frequency axis, $\Sigma(\mathbf{k,\nu})=\Re\Sigma(\mathbf{k},i\nu\rightarrow 0)+i\Im\Sigma(\mathbf{k},i\nu\rightarrow 0)+(1-1/Z_{\mathbf{k}})\nu+\mathcal{O}(\nu^2)$, a dichotomy in how non-mean-field fluctuations manifest themselves in the self-energy can be noticed:
Among the expansion coefficient, only the static real-part $\Re\Sigma(\mathbf{k},0)$ acquires a notable momentum-dependence in Fig.~\ref{fig:ADGAresults}. The coefficient that accounts, to first order, for the dynamical dependency of the self-energy---the quasi-particle weight $Z_\mathbf{k}$---is virtually independent of momentum: $Z_\mathbf{k}=Z$. As a consequence, the self-energy can be decomposed into $\Sigma(\mathbf{k},\nu)=\Sigma^{static}(\mathbf{k})+\Sigma^{local}(\nu)$. This ``space-time separation'' of correlation effects in SrVO$_3$ is congruent with earlier results from many-body perturbation theory\cite{jmt_pnict,Tomczak2014,jmt_sces14}, and also with D$\Gamma$A results for the Hubbard model in three dimensions\cite{Schaefer2015}. This empirical finding might pave the way for realizing more efficient many-body methods (for a space-time-separated {\it GW} scheme see Ref.~\citen{Schaefer2015}).

\section{Conclusions and outlook}
\label{Sec:conclusion}

In conclusion, we have reviewed the latest advances in 
diagrammatic extensions of DMFT for {\it ab initio} electronic structure calculations. The basic approximation of AbinitioD$\Gamma$A   is to take,
as the irreducible vertex $\Gamma$,
 the  Coulomb interaction and all local vertex corrections.
Solving the Bethe-Salpeter equation then yields the full vertex 
$F$ which includes, among others, non-local 
spin fluctuations and screening. The same kind of contributions are also taken into account for 
the transversal particle-hole channel which is related by crossing symmetry.
From $F$ in turn, the AbinitioD$\Gamma$A self-energy is obtained through the  Schwinger-Dyson equation of motion. These four steps are summarized in the flow diagram  Fig.~\ref{fig:flow}.

If only a few orbitals are considered, calculating the local vertex corrections  is numerically the most expensive  part. We have hence  reviewed recent improvements in CT-QMC for calculating local two-particle vertex functions. This involves worm sampling and makes use of the vertex asymptotic for high frequencies.

\new{Similar realistic, multi-orbital calculations should also be possible in the future using other
diagrammatic extensions of DMFT such as the dual fermion 
\cite{Rubtsov2008} or the one-particle irreducible approach \cite{Rohringer2013}.
For local interactions,  all these approaches are closely related
and  differ only in which vertex is taken (irreducible or full) and by which Green function lines these building-blocks are connected, see \cite{VertexRMP} for an overview. We have seen here that the D$\Gamma$A can be easily extended to non-local interactions. As it builds upon the irreducible vertex, relevant diagrams such as those of $GW$ are naturally included. %considered.
In the case of the  dual fermion approach which is based on the full vertex, more extensive modifications are necessary such as introducing additional bosonic dual  variables.}

As an application of AbinitioD$\Gamma$A we have presented results for the correlated metal SrVO$_3$.
These highlight the importance of non-local
self-energies even for fairly isotropic systems that are far from any ordering instabilities.
Much larger effects can be expected for materials that are anisotropic and/or that are in the vicinity of second order phase transitions.
The here reviewed AbinitioD$\Gamma$A approach presents a promising route towards describing these systems
and elucidating the microscopic fabric of non-local fluctuations of charge and spin, e.g., in the context of (quantum) criticality.

\medskip

\paragraph{Acknowledgments}

%\acknowledgment

We thank   A.\ Katanin, J.\ Kune\v{s}, G.\ Li, G.\ Rohringer, T.\ Sch\"afer, D.\ Springer, A.\ Toschi,  M.\ Wallerberger, and N.\ Wentzell,  for useful discussions. This work has been supported  by the  European Research Council under the European Union's Seventh
Framework Program (FP/2007-2013) through  ERC grant agreement n.\ 306447;
AG also thanks the Doctoral School W1243 Solids4Fun (Building Solids for Function) of the Austrian Science Fund (FWF) and PG the  Vienna Scientific Cluster (VSC) Research Center funded by
the Austrian Federal Ministry of Science, Research and Economy  (bmwfw). The  computational  results presented have been achieved using the VSC.

%\begin{verbatim}
%\profile{Taro Butsuri}{was born in Tokyo, Japan in 1965. ...}
%\end{verbatim}

%\bibliographystyle{jpsj}
%\bibliography{main,addrefs}

\end{document}